\documentclass[fleqn,usenatbib]{mnras}

\usepackage{newtxtext,newtxmath}

\usepackage[T1]{fontenc}

\DeclareRobustCommand{\VAN}[3]{#2}
\let\VANthebibliography\thebibliography
\def\thebibliography{\DeclareRobustCommand{\VAN}[3]{##3}\VANthebibliography}

\usepackage{graphicx}	
\usepackage{amsmath}	

\newcommand{\appropto}{\mathrel{\vcenter{
  \offinterlineskip\halign{\hfil$##$\cr
    \propto\cr\noalign{\kern2pt}\sim\cr\noalign{\kern-2pt}}}}}

\title[Magnetars and Late-Time Infrared]{Late-Time Infrared Cooling in Magnetar-Driven Supernovae}

\author[Omand]{
Conor M. B. Omand$^{1}$\thanks{E-mail: c.m.omand@ljmu.ac.uk} \href{https://orcid.org/0000-0002-9646-8710}{\includegraphics[scale=0.5]{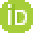}}
\\
$^{1}$Astrophysics Research Institute, Liverpool John Moores University, Liverpool Science Park IC2, 146 Brownlow Hill, Liverpool, UK, L3 5R}

\date{Accepted XXX. Received YYY; in original form ZZZ}

\pubyear{2025}

\begin{document}
\label{firstpage}
\pagerange{\pageref{firstpage}--\pageref{lastpage}}
\maketitle

\begin{abstract}
A central magnetar engine is commonly invoked to explain energetic supernovae, which should have multiple signals in multiwavelength emission.  Photoionization from the pulsar wind nebula (PWN) can create distinct spectroscopic signals in the nebular phase.  Recent models suggest infrared emission, particularly from Ne II, can be prominent at late times.  This work examines the cooling power of optical and infrared transitions to determine which lines contribute strongly to cooling and on what timescale.  The models show infrared cooling becomes strong at $\sim$ 3 years post-explosion and dominates by 6 years, with [Ne II] 12.8$\mu$m being the strongest coolant.  The fraction of total cooling in the infrared increases sharply once the PWN luminosity decreases below 10$^{40}$ erg s$^{-1}$, and this fraction also increases with increasing ejecta mass and decreasing average PWN photon energy.  However, the emission from [Ne II] 12.8$\mu$m increases with increasing PWN luminosity and increasing ejecta mass.  Cooling at 1 year is dominated by optical O and S lines, with infrared Ar, Ni, and Ne lines becoming strong at 3 years.  Optical cooling is almost negligible at 6 years, with the supernova cooling almost entirely through mid- and far-infrared transitions.  JWST spectroscopy with MIRI should be able to detect these lines out to $z \sim 0.1$.  Supernovae with higher magnetic fields transition to infrared cooling on earlier timescales, while infrared-dominated supernovae should have strong emission from neutral atoms and emit strongly in radio at sub-decade timescales.
\end{abstract}


\section{Introduction} \label{sec:intro}

The energetics of typical stripped-envelope core collapse supernovae are fairly well understood.  The kinetic energy of the ejecta, supplied by the supernova shock re-energized by a combination of neutrino emission from the proto-neutron star and multidimensional hydrodynamic instabilities, is generally around 10$^{51}$ erg \citep{Janka2012}.  The radiated energy of $\sim$ 10$^{49}$ erg is supplied by the $\sim$ 0.1 $M_\odot$ of radioactive $^{56}$Ni synthesized during the explosion \citep{Arnett1982}.  However, several classes of supernova, such as superluminous supernovae (SLSNe) and broad-lined Type Ic supernovae (SNe Ic-BL), have higher energies that cannot be explained under this paradigm \citep[e.g.][]{Gal-Yam2012, Taddia2019, Modjaz2019, Nicholl2021}.  Some models for explaining these supernovae include explosions of extremely massive stars in pair-instability (PISNe) or pulsational pair-instability supernovae \citep[PPISNe;][]{Barkat1967, Heger2002, Heger2003, Woosley2017, Schulze2024}; interaction with circumstellar material \citep[CSM;][]{Chatzopoulos2012, Chatzopoulos2013, Villar2017, Jiang2020, Margutti2023} from steady winds, eruptive mass loss, or binary stripping \citep{Smith2014}; and energy injection from fallback accretion onto a newborn black hole or neutron star \citep{Dexter2013, Moriya2018}, interaction with a jet \citep{MacFadyen1999, Gottlieb2024}, or the spin-down of a highly magnetized, rapidly rotating neutron star (magnetar) \citep{Kasen2010, Woosley2010}.

As the magnetar spins-down, the rotational energy from the magnetar is converted into a relativistic wind.  This wind interacts with the ejecta, sending a forward shock into the ejecta and a reverse shock back into the wind.  This shocked wind, known as a pulsar wind nebula (PWN) heats and accelerates the ejecta \citep{Kasen2010, Woosley2010, Omand2024}, and emits non-thermal emission from synchrotron radiation and inverse Compton scattering \citep{Gaensler2006}.  The PWN can be detected in radio, X-ray, and gamma-ray wavelengths \citep{Kotera2013, Murase2015, Kashiyama2016, Omand2018, Eftekhari2021, Murase2021, Omand2025grb}, and has been detected at $\sim$ a decade post-explosion in PTF10hgi and SN2012au \citep{Eftekhari2019, Law2019, Mondal2020, Hatsukade2021, Stroh2021}.  Dust formed in the supernova ejecta can be heated by the PWN, producing bright continuum emission in the infrared \citep{Omand2019, Chen2021, Sun2022irslsn}.  Hydrodynamic instabilities from the PWN/ejecta interaction can shred the ejecta \citep{Chen2016, Chen2020, Blondin2017, Suzuki2017, Suzuki2019, Suzuki2021, Eiden2025}, producing clumpy filamentary structures similar to those seen in the Crab Nebula \citep{Clark1983, Bietenholz1991, Temim2006, Temim2024, Omand2025Crab}.  The asymmetry from the formation of these structures may cause a polarization signal similar to those observed in several potential magnetar-driven supernovae \citep{Leloudas2015, Leloudas2017, Inserra2016, Saito2020, Poidevin2022, Poidevin2023, Poidevin2025, Pursiainen2022, Pursiainen2023}.  Magnetars have also been invoked to explain some long gamma-ray bursts (GRBs) and their associated supernovae \citep[e.g.][]{Usov1992, Metzger2011, Kumar2024, Srinivasaragavan2024, RomanAguilar2025}.

Another signature of a magnetar engine is the presence of lines from highly ionized species, like O III, in the nebular spectrum of the supernova \citep{Chevalier1992, Dessart2012, Dessart2019, Dessart2024, Jerkstrand2017, Omand2023}, which makes spectroscopy a powerful tool for diagnosing the power source of luminous supernovae.   These lines have been seen in several oxygen-rich supernova remnants \citep{Kravtsov2025} as well six years post-explosion in SN 2012au \citep{Milisavljevic2018}.  Modelling of the nebular spectrum of SN 2012au produced results that were also consistent with light curve and radio modelling \citep{Stroh2021, Omand2023}, leading to the most comprehensive characterization of the magnetar-driven supernova to date.

Both \citet{Omand2023} and \citet{Dessart2024} focused primarily on optical emission, with both noting the prevalence of infrared emission, particularly from [Ne II] 12.8$\mu$m, at late times.  Both studies found that over 50\% of the line cooling could come from [Ne II] 12.8$\mu$m at late times in some models, but neither quantified the timescale for when infrared emission begins to dominate, nor how ubiquitous this behaviour is among magnetar-driven supernovae.  With the early success of the James Webb Space Telescope (JWST), understanding the ubiquity and timescales from this emission is paramount.  This study examines the cooling rates of the models from \citet{Omand2023} to examine the prevalence of infrared cooling for grids of models at different timescales to help identify when certain lines emerge and if the emergence of infrared lines and luminosity of the lines are correlated with any ejecta properties, allowing for extrapolation outside the model grid.  It is worth noting that cooling rates alone are not suitable for quantitatively approximating spectra, sometimes differing from the predicted line luminosity by over an order of magnitude, so this data is only suitable for predicting/identifying lines and for qualitative trends.  Section \ref{sec:trans} overviews the model grids and cooling rate lists, Section \ref{sec:results} presents the analysis of the model grid, Section \ref{sec:disc} presents the discussion, and Section \ref{sec:conc} concludes.

\section{Model Data} \label{sec:trans}

\citet{Omand2023} presented radiative transfer simulation grids using the non-local thermodynamic equilibrium (non-LTE) \textsc{SUMO} spectral synthesis code \citep{Jerkstrand2011, Jerkstrand2012} modified to allow photon injection from the inner boundary, which was used to approximate the photons injected from the PWN.  These radiative transfer models were presented for two epochs, 1 year and 6 years post-explosion, and for two ejecta compositions, a pure oxygen composition and a realistic but fully mixed Type Ibc composition from a 25 $M_\odot$ ZAMS mass progenitor model from \citet{Woosley2007}.  The model grid used three different parameters: the ejecta mass ($M_{\rm ej}$), the PWN luminosity at that that epoch ($L_{\rm PWN} (t)$, subsequently referred to as $L_{\rm PWN}$), and the PWN spectral peak parameter ($T_{\rm PWN}$).  The model grids have parameter ranges from $10^5 - 10^6$ K for $T_{\rm PWN}$ at all epochs, have different ranges for $L_{\rm PWN}$ at each epoch roughly corresponding to initial dipole magnetic field of $10^{14} - 10^{15}$ G, and have $M_{\rm ej}$ ranging from $1 - 10$ $M_\odot$ at 6 years and $1.5 - 4$ $M_\odot$ at 1 and 3 years.  The models all contain a single zone with inner and outer velocities of 2000 and 3000 km s$^{-1}$, respectively.  These choices were motivated by the density profiles and mixing properties found in numerical simulations of magnetar-driven supernovae \citep{Chen2020, Suzuki2021}.  All other assumptions are noted in Section 2 of \citet{Omand2023}.

\begin{figure}
    \centering
    \includegraphics[width=\linewidth]{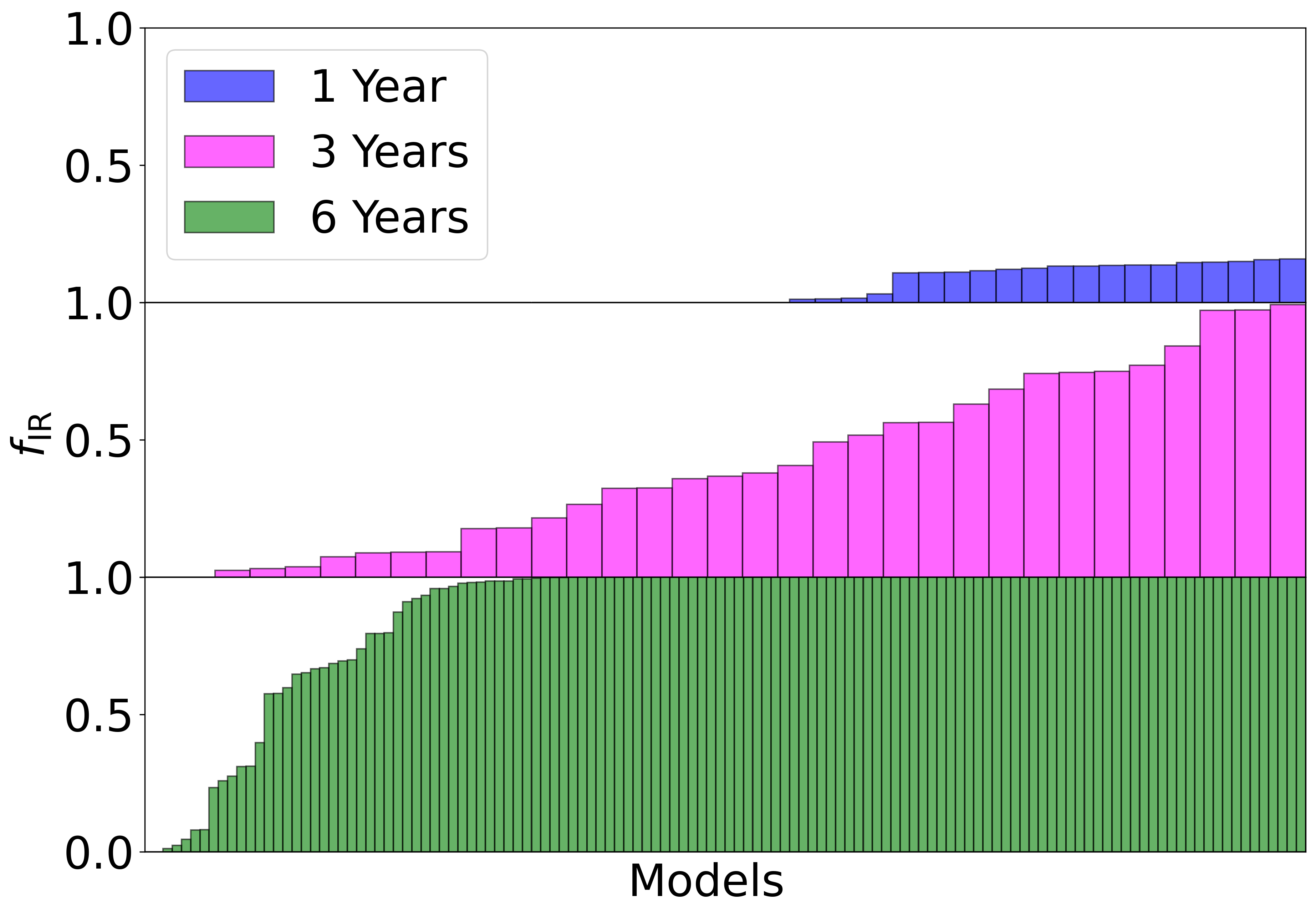} \\
    \includegraphics[width=\linewidth]{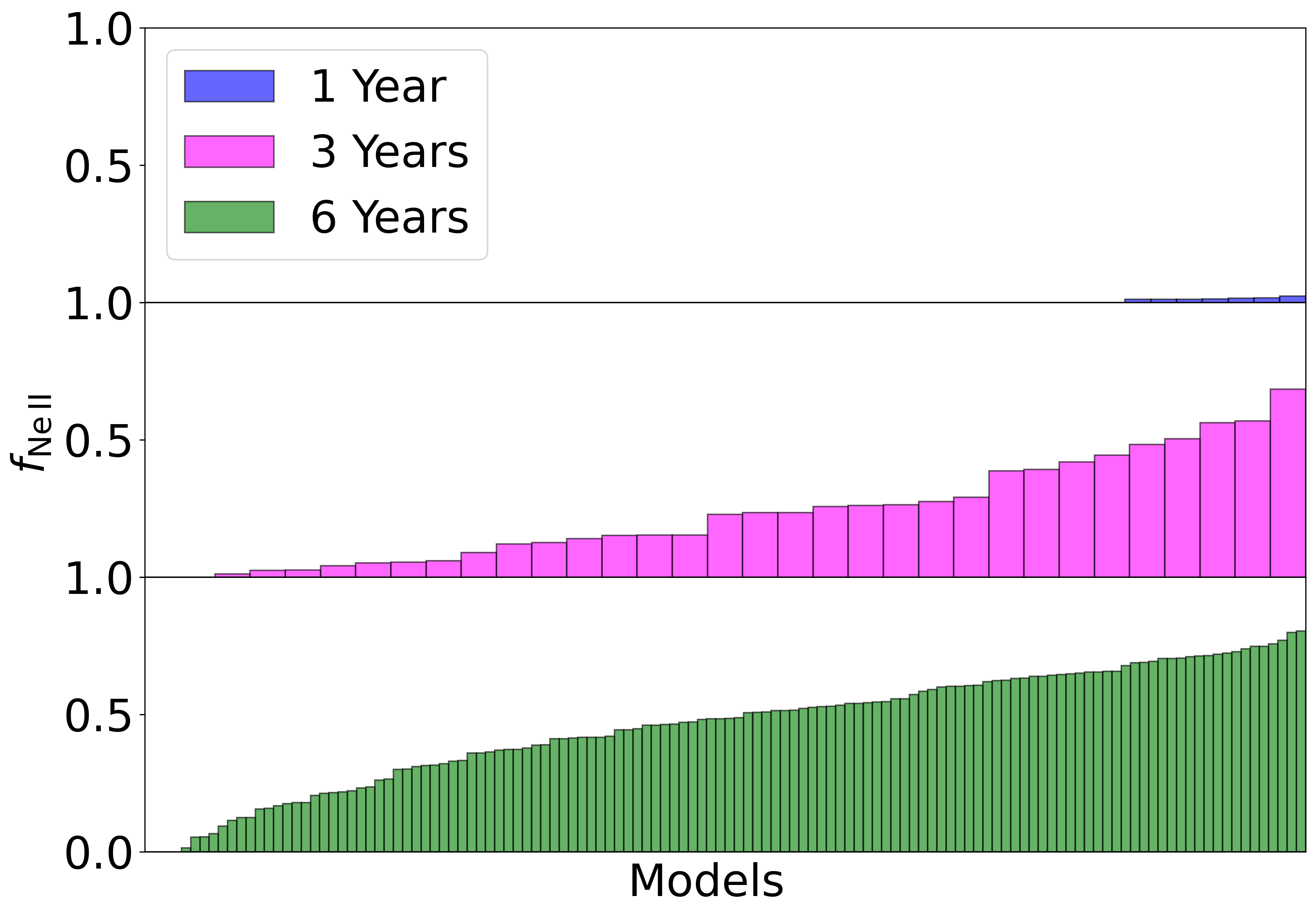}
    \caption{The total power of cooling transitions in the infrared ($> $ 1 $\mu$m, top) and though [Ne II] 12.8$\mu$m alone (bottom) as a fraction of the total power of cooling transitions at all wavelengths.  The three panels in each figure represent three epochs: 1 year ($N = $ 45, top), 3 years ($N = $ 33, middle), and 6 years ($N = $ 126, bottom).  Each bar represents a different model from the grids at each epoch, and are sorted according to the highest values of $f_{\rm IR}$ and $f_{\rm Ne \,II}$.}
    \label{fig:cooling}
\end{figure}

Two public lists of cooling transitions\footnote{\url{https://github.com/conoromand/LineCooling} \label{fn:link}} were created from the output of the Type Ibc models at each epoch, as well as a previously unpublished model grid at 3 years that had been used for model calibration.  The model grids include $N = $ 45 models at $t$ = 1 year, $N = $ 33 models at $t$ = 3 years, and $N = $ 126 models at $t$ = 6 years.  Radiative transfer was only performed up to 10 $\mu$m, and is thus missing many key mid-infrared and far-infrared lines, but cooling rates were calculated for transitions at all wavelengths.  For each model, each transition between 0.1 and 100 $\mu$m is recorded with its total power in that epoch.  One list at each epoch contains all cooling transitions, while the other contains only transitions
with $>$ 1\% of the total cooling power for that model at that epoch.  This threshold is implemented to examine only the most dominant coolants in each model, and the analyses within the paper are based on the lists with the thresholds, although this should only affect ubiquity statistics. The lists with thresholds contain a total of 52 different transitions (28 optical/UV, 24 infrared) from 24 ions of 11 elements.  I reiterate that cooling rates alone are only suitable for predicting/identifying lines and for qualitative trends, and this study is not attempting to make predictions for line luminosities.

\section{Model Grid Analysis} \label{sec:results}

The infrared and [Ne II] 12.8$\mu$m cooling fractions $f_{\rm IR}$ and $f_{\rm Ne \,II}$, which denote the fraction of total power from the cooling transitions $\dot{E}_{\rm cool}$ through infrared transitions and the [Ne II] 12.8$\mu$m transition alone, are defined as

\begin{align}
    f_{\rm IR} =& \frac{\sum_{i, \lambda > 1 \mu {\rm m}}\dot{E}_{\mathrm{cool},\,i}}{\sum_{i}\dot{E}_{\mathrm{cool},\,i}} \label{eqn:fir}\\
    f_{\rm Ne \,II} =& \frac{\dot{E}_{\mathrm{cool},\,[\mathrm{Ne \,II}]\,12.8\,\mu\mathrm{m}}}{\sum_{i}\dot{E}_{\mathrm{cool},\,i}} \label{eqn:fNe}
\end{align}
where $i$ denotes the different transitions and $\dot{E}_{\mathrm{cool},\,i}$ denotes the cooling power through the $i$ transition.  Figure \ref{fig:cooling} shows the $f_{\rm IR}$ and $f_{\rm Ne \,II}$ for each model and each epoch, sorted by the $f_{\rm IR}$ and $f_{\rm Ne \,II}$ values.  At 1 year no models have more than $\sim$ 20\% of cooling through infrared transitions with only a small amount coming through the [Ne II] 12.8$\mu$m transition.  At 3 years, the models show a diversity of behaviour, with as many models showing a small fraction of transition power in the infrared as there are models dominated by transitions in the infrared.  Some models show up to $\sim$ 60\% of the total cooling power through the [Ne II] 12.8$\mu$m, and models generally show about 50\% of the infrared cooling power coming through the [Ne II] 12.8$\mu$m transition.  At 6 years, most models are cooled primarily through infrared transitions, with $\sim$ 75\% of the models showing a negligible amount of cooling through optical/UV transitions.  Some models show up to 80\% of cooling through only [Ne II] 12.8$\mu$m, with the majority of models cooling $\gtrsim$ 50\% through this one transition.

\begin{figure}
    \centering
    \includegraphics[width=0.99\linewidth]{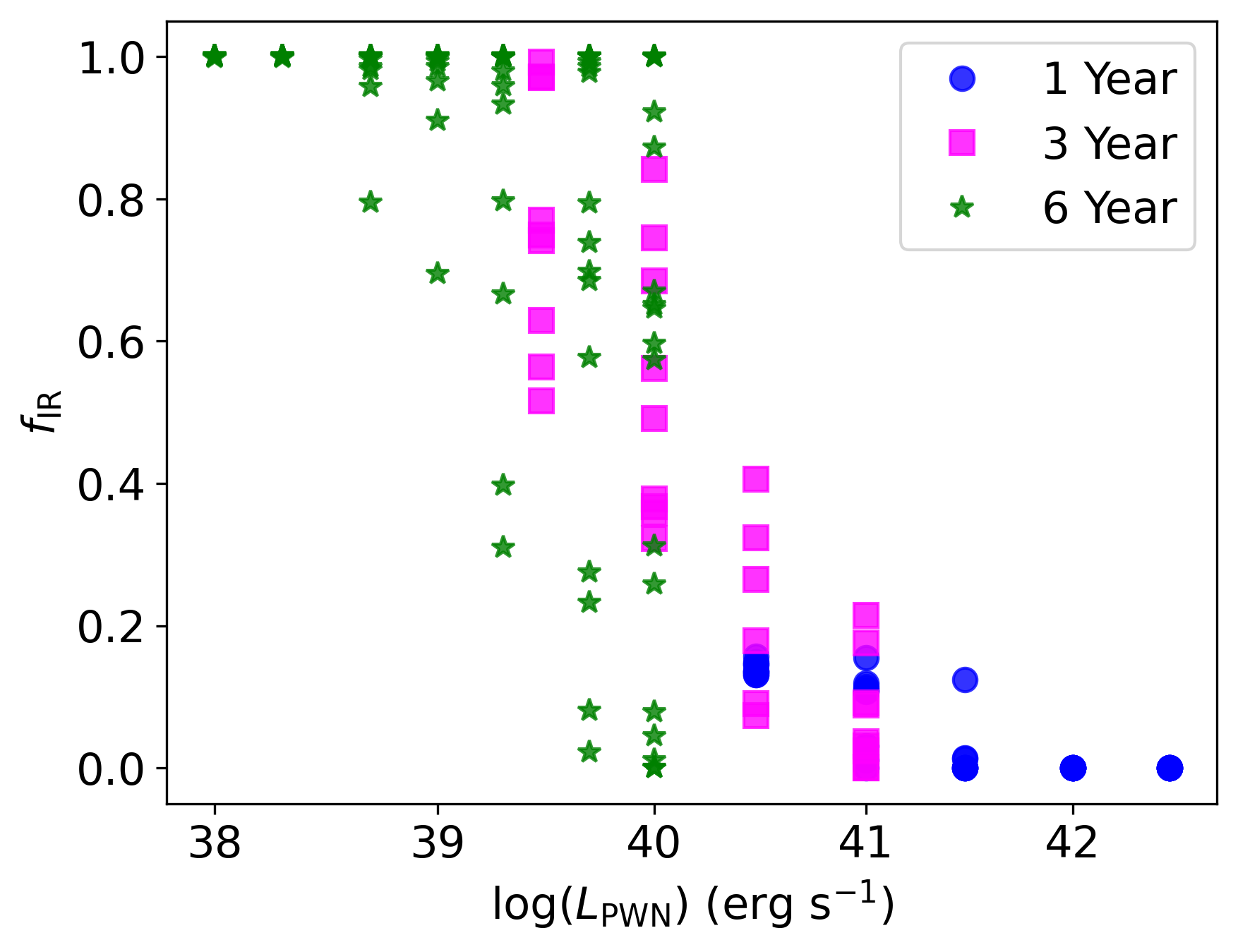} \\
    \includegraphics[width=0.99\linewidth]{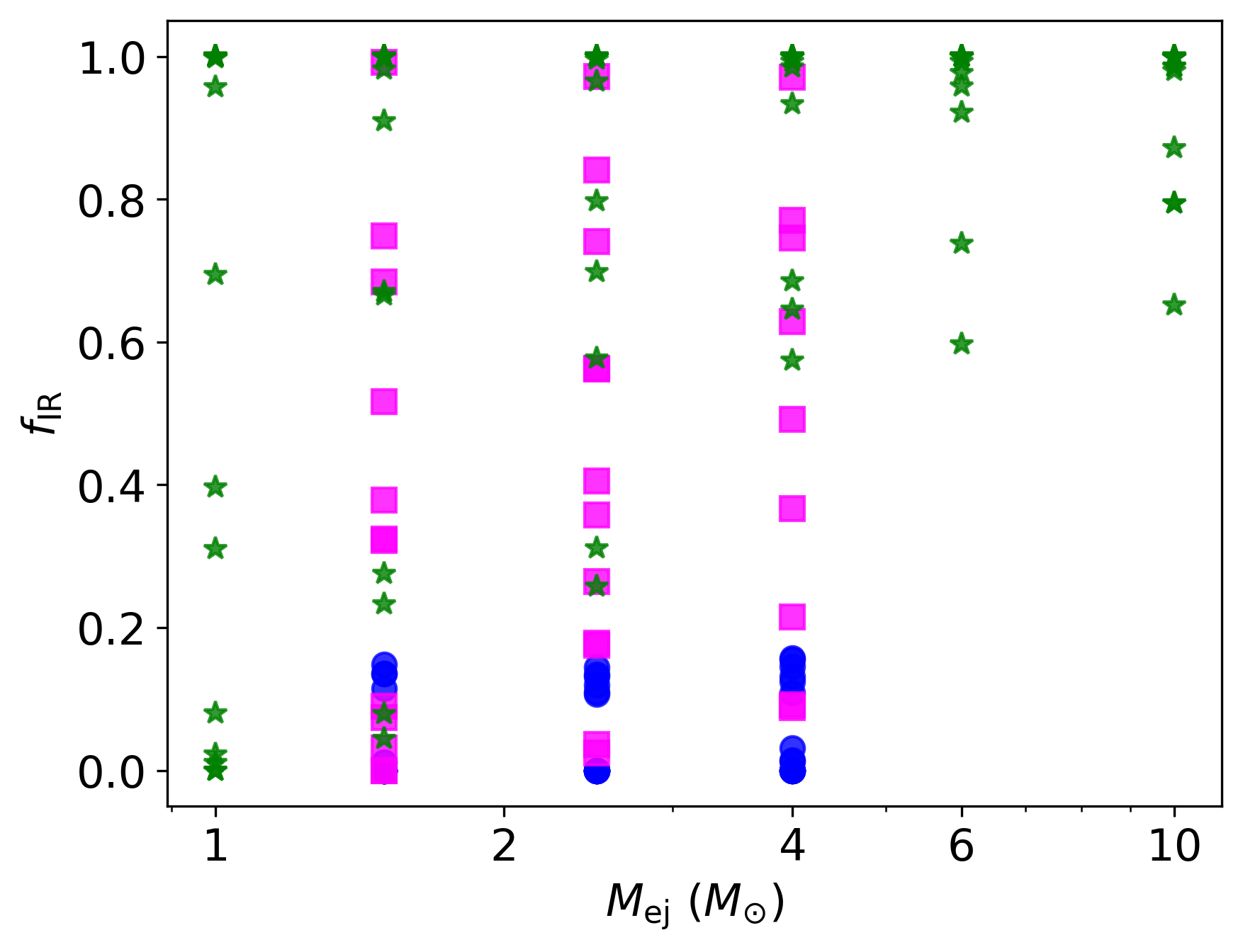} \\
    \includegraphics[width=0.99\linewidth]{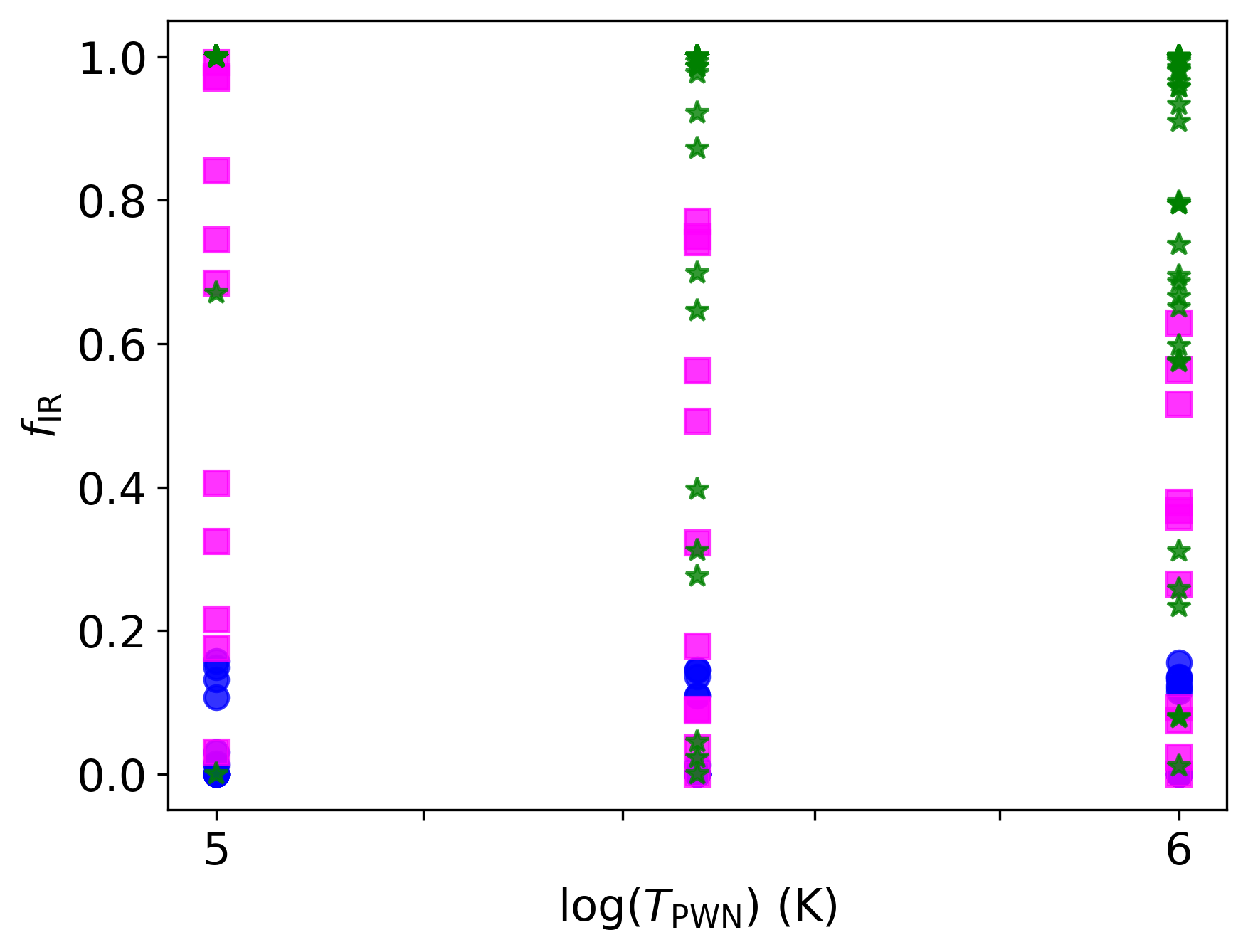}
    \caption{The fraction of cooling power in the infrared $f_{\rm IR}$ for all models as a function of $L_{\rm PWN}$ (top), $M_{\rm ej}$ (middle), and $T_{\rm PWN}$ (bottom).}
    \label{fig:corr}
\end{figure}

Figure \ref{fig:corr} shows $f_{\rm IR}$ for all models as a function of $L_{\rm PWN}$, $M_{\rm ej}$, and $T_{\rm PWN}$.  $f_{\rm IR}$ as a function of $L_{\rm PWN}$ shows little dependence on the epoch of the model, with $f_{\rm IR}$ $\sim$ 0 at $L_{\rm PWN} \gtrsim 10^{41}$ erg s$^{-1}$, $f_{\rm IR}$ $\sim$ 1 at $L_{\rm PWN} \lesssim 10^{39}$ erg s$^{-1}$, and lots of variation in the models at more intermediate values of $L_{\rm PWN}$.  The time dependence of $f_{\rm IR}$ in the models seems mostly due to how $L_{\rm PWN}$ scales with time.  $f_{\rm IR}$ is also correlated with either $M_{\rm ej}$ and $T_{\rm PWN}$, with an increase in $f_{\rm IR}$ for increasing values of $M_{\rm ej}$ and decreasing values of $T_{\rm PWN}$ at later times, although this is not as strong as the correlation with $L_{\rm PWN}$.  These correlations also hold for $f_{\rm Ne \,II}$ at 3 and 6 years, although the fraction of power in that transition at 1 year is too low to see if the correlations hold at that epoch.

\begin{figure}
    \centering
    \includegraphics[width=0.99\linewidth]{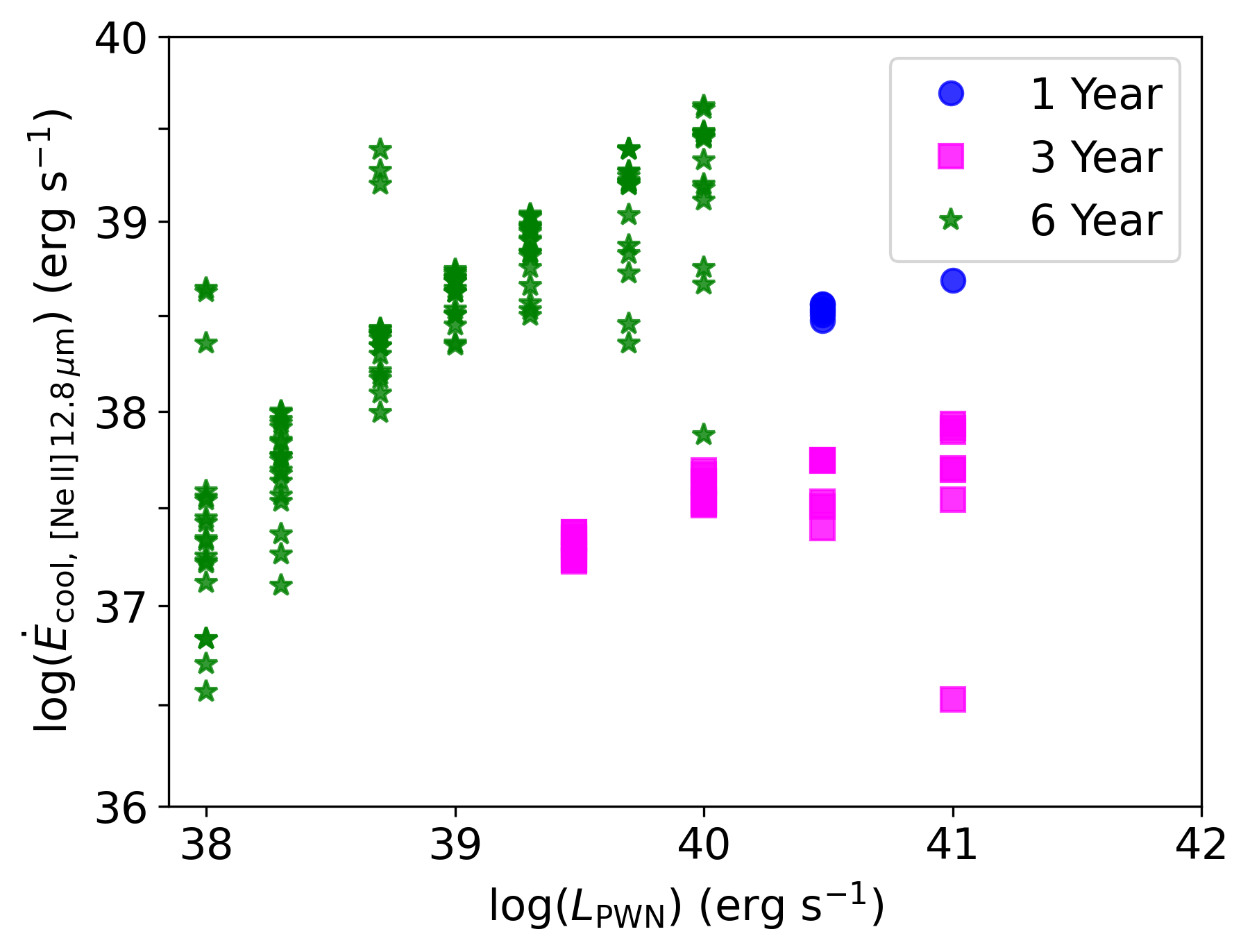} \\
    \includegraphics[width=0.99\linewidth]{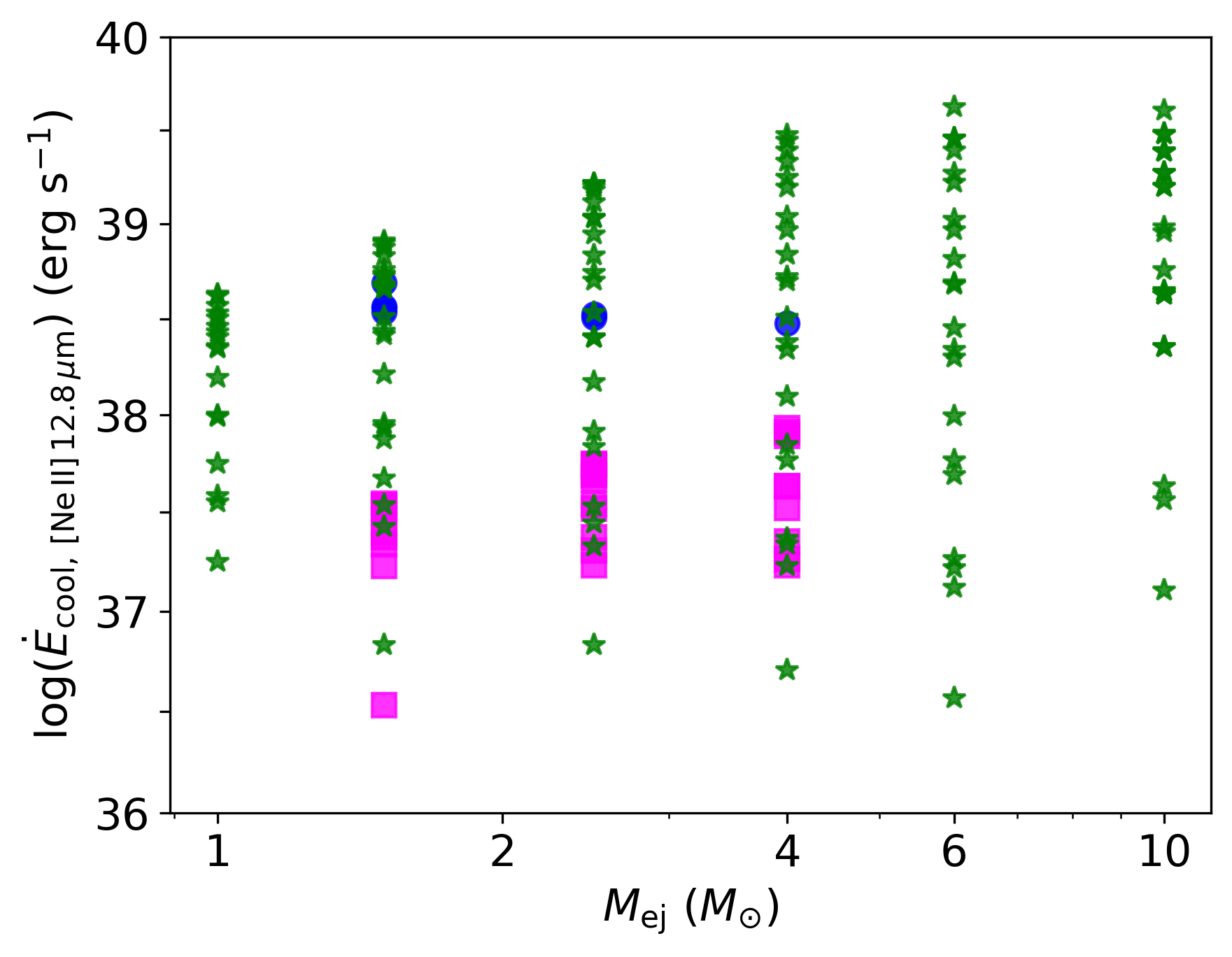} \\
    \includegraphics[width=0.99\linewidth]{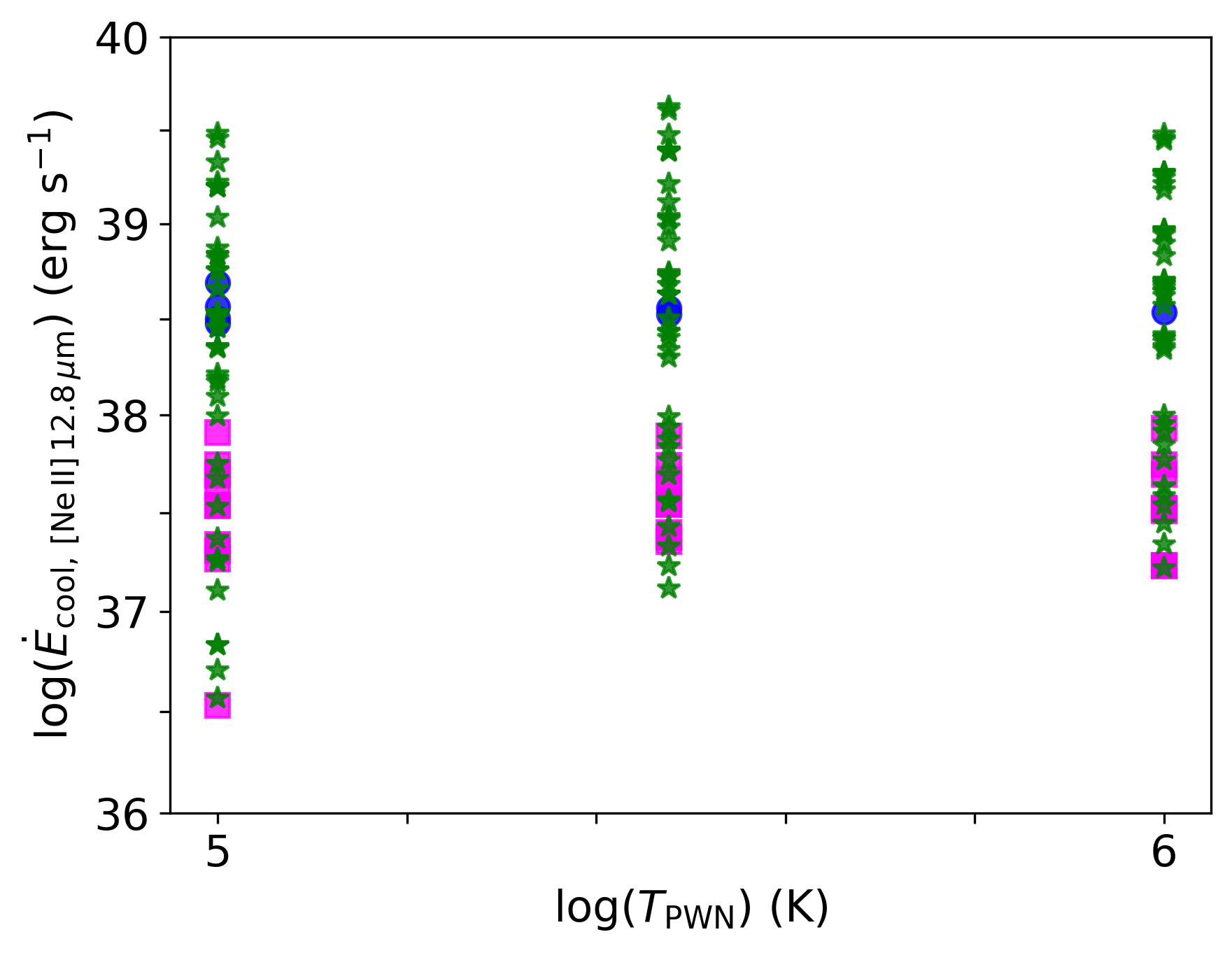}
    \caption{The cooling power of [Ne II] 12.8$\mu$m for all models as a function of $L_{\rm PWN}$ (top), $M_{\rm ej}$ (middle), and $T_{\rm PWN}$ (bottom).}
    \label{fig:ne_corr}
\end{figure}

The detectability of certain lines, such as [Ne II] 12.8$\mu$m, depends more on their values of $\dot{E}_{\rm cool}$ than on $f_{\rm IR}$.  Figure \ref{fig:ne_corr} shows $\dot{E}_{\mathrm{cool},\,[\mathrm{Ne II}]\,12.8\,\mu\mathrm{m}}$ for all models as a function of $L_{\rm PWN}$, $M_{\rm ej}$, and $T_{\rm PWN}$.  $\dot{E}_{\rm cool}$ shows the opposite correlation with $L_{\rm PWN}$ compared to $f_{\rm IR}$, increasing with increasing $L_{\rm PWN}$ at each epoch.  This shows that both optical and infrared luminosities are decreasing as $L_{\rm PWN}$ decreases, but infrared cooling decreases slower, causing the total infrared cooling fraction to increase.  The behavior of $\dot{E}_{\rm cool}$ with time shows an initial decrease between 1 and 3 years before increasing again at 6 years, although many models at 1 year do not show [Ne II] 12.8$\mu$m emission, so this is likely a selection effect.  $\dot{E}_{\mathrm{cool},\,[\mathrm{Ne II}]\,12.8\,\mu\mathrm{m}}$ shows an increase as $M_{\rm ej}$ increases at the later two epochs, and shows no significant correlation with $T_{\rm PWN}$.

\begin{figure}
    \centering
    \includegraphics[width=0.99\linewidth]{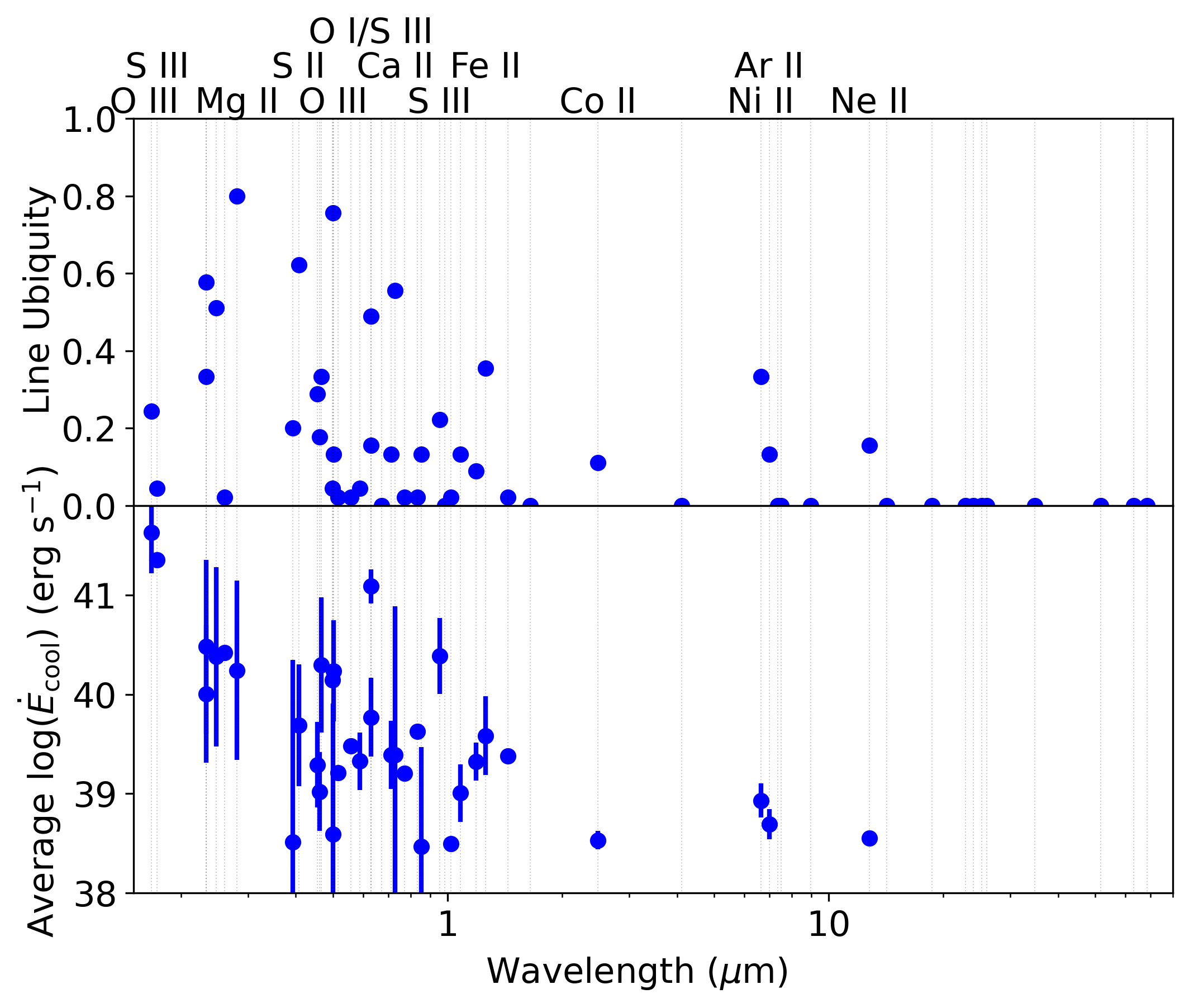} \\
    \includegraphics[width=0.99\linewidth]{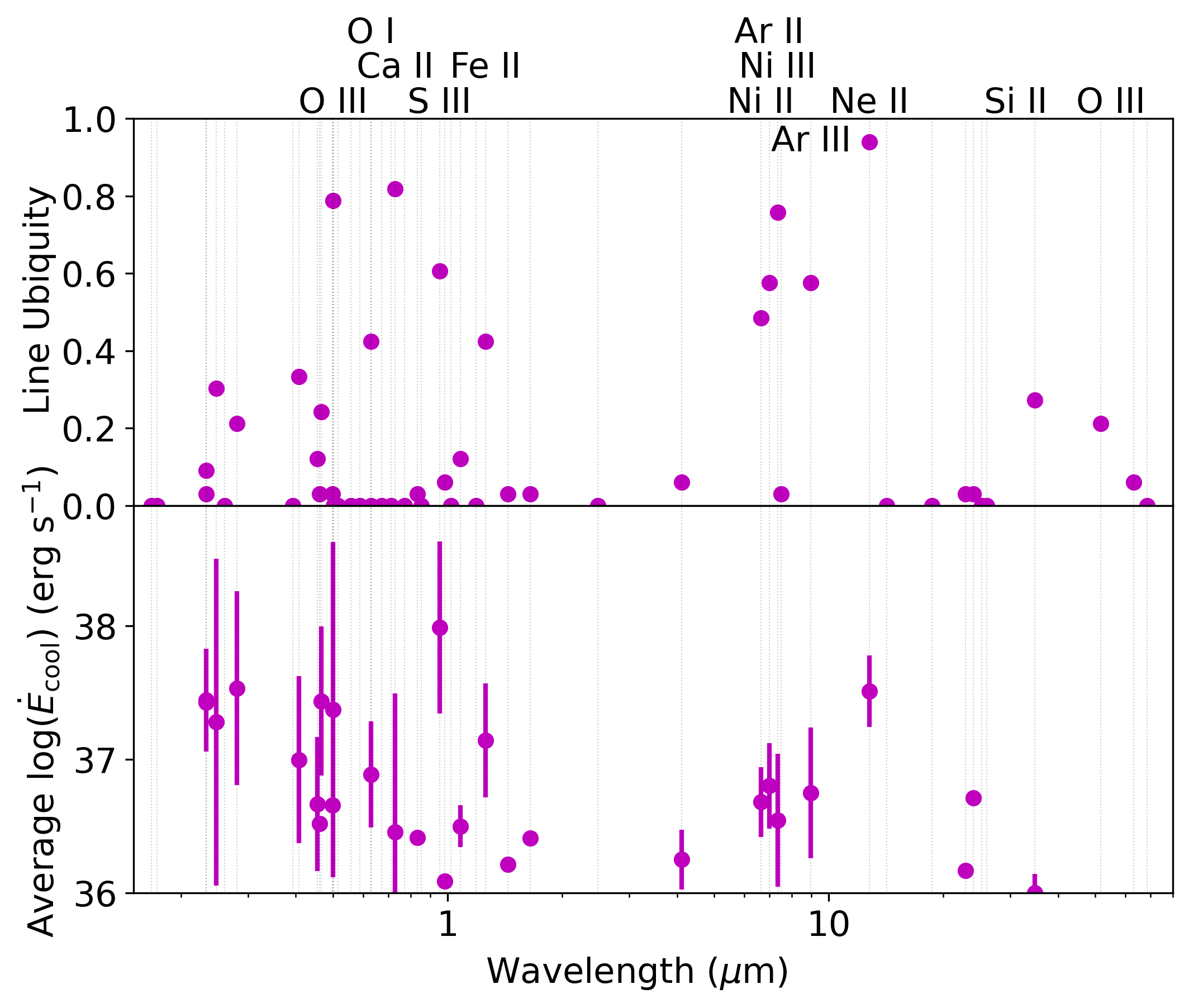} \\
    \includegraphics[width=0.99\linewidth]{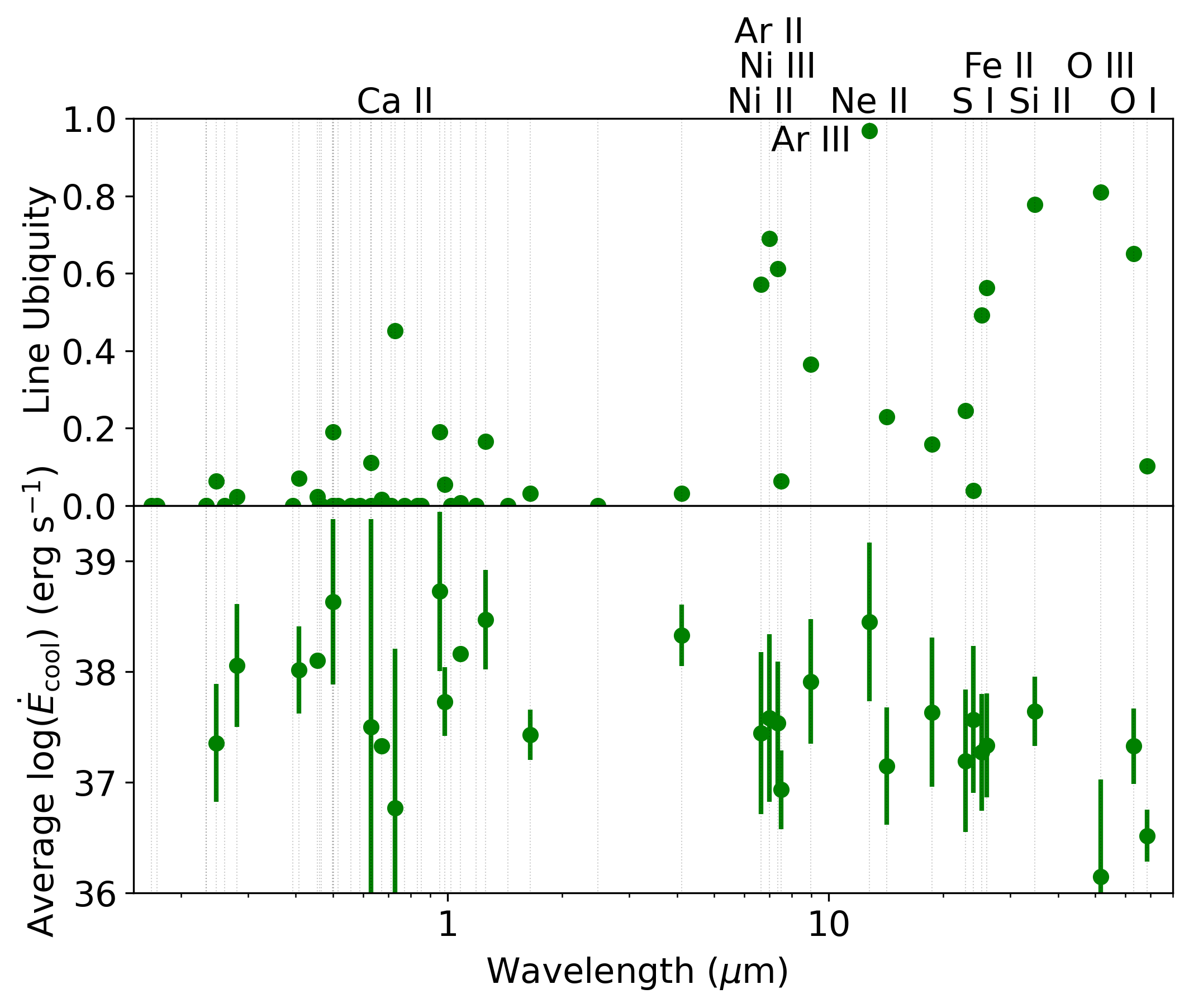}
    \caption{The fraction of models a particular transition is listed, referred to as the line ubiquity, at each 1 year ($N = $ 45, top), 3 years ($N = $ 33, middle), and 6 years ($N = $ 126, bottom); as well as the average $\dot{E}_{\rm cool}$ for each transition at that epoch.  The ions for prominent transitions are labeled above.}
    \label{fig:ubps}
\end{figure}

Although [Ne II] 12.8$\mu$m is the strongest coolant in most late-time cases, there are other prominent infrared coolants at all epochs.  Figure \ref{fig:ubps} shows the fraction of models for which a particular transition is listed, referred to as the line ubiquity, at each epoch, as well as the average $\dot{E}_{\rm cool}$ for each transition at that epoch.  $\dot{E}_{\rm cool}$ is averaged in log-space, and the averaging is only done for the models where that transition is present.  Line ubiquity is sensitive to the choice of 1\% the cooling power threshold, so line ubiquity here represents the fraction of models where the transition is one of the dominant coolants, not the fraction of models where the transition cools at all.  At 1 year,  the most ubiquitous lines are in the optical and UV, particularly Mg II $\lambda\lambda$ 2795.5, 2802.7; [O III] $\lambda\lambda$ 4958.9, 5006.8;  [S II] $\lambda\lambda$ 6716.4, 6730.8;  and [Ca II] $\lambda\lambda$ 7291.5, 7323.9. Several optical/UV lines are not ubiquitous, but are strong coolants when they are present, including [O III] $\lambda\lambda$ 1660.8, 1666.2; [S III] $\lambda\lambda$ 1713.1, 1728.9; [S III] $\lambda$ 6312.1; and [S III] $\lambda$ 9530.6.  In infrared, [Fe II] 1.26$\mu$m, [Ni II] 6.634$\mu$m, and [Ne II] 12.81$\mu$m are present in about 20 -- 30\% of models, but are not prominent coolants compared to the optical transitions.

At three years, both optical and infrared lines become ubiquitous.  In the optical, [O III] $\lambda\lambda$ 4958.9, 5006.8;  [O I] $\lambda\lambda$ 6300.3, 6363.8; [Ca II] $\lambda\lambda$ 7291.5, 7323.9; and [S III] $\lambda$ 9530.6 are in $\gtrsim$ 50\% of models, with [S III] $\lambda$ 9530.6 being the strongest coolant.  In infrared, [Fe II] 1.26$\mu$m, [Ni II] 6.634$\mu$m, [Ar II] 6.983$\mu$m, [Ni III] 7.347$\mu$m, [Ar III] 8.991$\mu$m, and [Ne II] 12.81$\mu$m are in $\gtrsim$ 50\% of models, with [Ne II] 12.81$\mu$m being the strongest infrared coolant by about an order of magnitude.  Several far-infrared transitions start to appear at this epoch, including [Si II] 34.81$\mu$m and [O III] 51.8$\mu$m, but they are only in $\sim$ 20\% of models and are weak coolants compared to the mid-infrared lines.

At six years, the only optical line present in $\gtrsim$ 50\% of the models is [Ca II] $\lambda\lambda$ 7291.5, 7323.9; which cools less than the majority of infrared lines.  The six year models again show ubiquitous emission from [Ni II] 6.634$\mu$m, [Ar II] 6.983$\mu$m, [Ni III] 7.347$\mu$m, [Ar III] 8.991$\mu$m, and [Ne II] 12.81$\mu$m, with all of these lines having higher $\dot{E}_{\rm cool}$ than at 3 years by up to an order of magnitude.  [Ne II] 12.81$\mu$m is again the strongest coolant among all the ubiquitous models.  Far-infrared transitions become much more prominent at this epoch, with [S I] 25.249$\mu$m, [Fe II] 25.981$\mu$m, [Si II] 34.81$\mu$m, [O III] 51.8$\mu$m, and [O I] 63.168$\mu$m appearing in $\gtrsim$ 50\% of models and having similar cooling power as the mid-infrared transitions.

\section{Discussion} \label{sec:disc}

The timescale for infrared transitions to dominate cooling can be less than 3 years in supernovae with high ejecta masses and low PWN luminosities.  Even with the variety of $f_{\rm IR}$ values at 3 years, the [Ne II] 12.81$\mu$m is the most ubiquitous transition in optical or infrared, and is the second strongest coolant on average behind only [S III] $\lambda$ 9530.6.  High ejecta masses and low PWN luminosities lead to an ejecta that has a low average ionization state.  \citet{Omand2023} showed that the dominant oxygen ion in this case in O I, with O II and O III ion fractions sometimes lower than 10$^{-3}$.  However, there do exist regions of parameter space where significant amounts of oxygen are in all of the first three ionization states, and this parameter region is likely where elements with high ionization energies, such as neon and argon, can exist in significant abundance at higher ionization states as well.  These states correspond to higher PWN luminosities within the epoch, which does correspond to higher $\dot{E}_{\rm cool}$ values for [Ne II] 12.81$\mu$m.  $T_{\rm PWN}$ does have a strong effect on the ionization state of the ejecta in a pure oxygen composition, resulting in higher average ionization for lower values of $T_{\rm PWN}$, but this effect is much weaker in a mixed composition.

Ejecta with low $L_{\rm PWN}$, high $M_{\rm ej}$, and low $T_{\rm PWN}$ has a much lower temperature than other regions of parameter space, with $T_{\rm ej} \sim$ 3000 K at 1 year and $T_{\rm ej} \sim$ 1000 K at 6 years \citep{Omand2023}.  Values of $L_{\rm PWN}$ at any specific epoch can be used to calculate the initial magnetic field $B$ of the magnetar (See Equation 5 from \citet{Omand2023}) if the magnetar is emitting as a pure vacuum dipole (see, e.g. \citet{Kashiyama2016, Lasky2017, Sarin2022} and \citet{Omand2024} non-dipole formalisms and emission properties), which is often inferred from light curve models like those implemented in \textsc{Mosfit} \citep{Guillochon2018} or \textsc{Redback} \citep{Sarin_redback}.  From Figure \ref{fig:corr}, the supernova undergoes a transition from optical- to infrared-dominated at $L_{\rm PWN} \sim 10^{40}$ erg s$^{-1}$, which translates to a timescale of

\begin{equation}
    t \sim 0.8 \text{ years } \left( \frac{B}{\text{10$^{15}$ G}}\right)^{-1} .
\end{equation}
In order to more easily probe infrared radiation in these supernovae, the inferred magnetic field should be lower than the average inferred from SLSNe \citep{Chen2023b, Gomez2024}.  The highest $L_{\rm PWN}$ value from the 6 year grid, $\sim 10^{40}$ erg s$^{-1}$, corresponds to a magnetic field of $\sim 10^{14}$ G, which is an average value for SLSNe.  The preferred large ejecta masses and low magnetic fields will have similar effects on the light curve width, with low magnetic fields leading to slower light curves due to lower acceleration of the ejecta \citep{Suzuki2021, Omand2024} and high ejecta masses leading to slower light curves due to the increased diffusion time.  Therefore, long-duration SLSNe, such as SN 2018ibb \citep{Schulze2024}, are excellent candidates for infrared spectroscopic observations to test the magnetar model.  The initial spin period does not factor into $L_{\rm PWN}$ at $t \gg t_{\rm SD}$, the spin-down time, so less-luminous supernovae powered by magnetars \citep[e.g.][]{Gomez2022} are still viable candidates for strong infrared emission.

The models from \citet{Omand2023} and the treatment $\dot{E}_{\rm cool}$ here have many caveats worth noting.  The linelists only have the most prominent cooling transition from each ion, meaning transitions from ions with multiple strong transitions will not be present in this data.  This is most notable for the Fe II and Fe III ions, which normally produce forests of line emission from 4000 $-$ 5500 \AA \, that is completely missing from these models.  The lack of radiative transfer here means the values of $\dot{E}_{\rm cool}$ can not be used as a proxy for line luminosities, limiting their quantitative usefulness.  These models neglect the formation of molecules \citep{Liljegren2020, Liljegren2023} and dust \citep{Nozawa2013, Omand2019}, which can cool the supernova significantly and emit in the near- and mid-infrared; such emission has already been detected in a suspected magnetar-driven SLSN, SN 2020wnt \citep{Tinyanont2023}.  The treatment of mixing and clumping in these models likely leads to an overabundance of cooling from S and Ca and an underabundance of cooling from O.  For more of a discussion on mixing and clumping in these models, see Section 6.2 from \citet{Omand2023}.

Given the many caveats of these results, comparing them with both observational and theoretical work is helpful to see if the results could hold predictive power.  Recent JWST observations of SN 1987A \citep{Fransson2024} and the Crab Nebula \citep{Temim2024} have seen emission lines from [Ar II] 6.983$\mu$m, [Ar III] 8.991$\mu$m, [Ne II] 12.81$\mu$m, and [Fe II] 25.981$\mu$m.  Late-time nebular spectra of SN 2012au show strong oxygen lines that are not ubiquitous in the 6-year models, although that could be due to the strong mixing between O and Ca in these models.  Models from \citet{Dessart2024} also predict that [Ni II] 6.634$\mu$m, [Ar II] 6.983$\mu$m, [Ni III] 7.347$\mu$m, [Ar III] 8.991$\mu$m, [Ne II] 12.81$\mu$m, [S I] 25.249$\mu$m, and [Fe II] 25.981$\mu$m will be strong coolants.  Figure 17 from \citet{Dessart2024} shows $f_{\rm Ne \,II}$ increasing initially but decreasing at later timescales.  While $f_{\rm Ne \,II}$ is peaking later than examined here for some progenitor models, this decrease is likely due to the constant $L_{\rm PWN}$ value from those models, rather than the more physical $L_{\rm PWN} \appropto t^{-2}$.  The $\dot{E}_{\mathrm{cool},\,[\mathrm{Ne II}]\,12.8\,\mu\mathrm{m}}$ values calculated in \citet{Dessart2024} are $\sim$ 10$^{38}$ erg s$^{-1}$, which is on the same order as the average values calculated here.

These model predictions and their consistency with observations of nearby objects imply that JWST spectroscopy will play a key role in identifying and characterizing magnetar-driven supernovae.  The redshift ranges for observing the [Ne II] 12.81$\mu$m line with the three channels for the Mid-infrared Instrument (MIRI) Medium-Resolution Spectrometer (MRS) \citep{Wells2015, Argyriou2023} are $z \lesssim 0.05$ for channel 3A, $0.05 \lesssim z \lesssim 0.22$ for channel 3B, and $0.22 \lesssim z \lesssim 0.41$ for channel 3C.  During these observations, channel 2 will be observing at around 7.5 $-$ 9.0 $\mu$m in the rest frame, meaning that it may be possible to observe either the [Ni III] 7.347$\mu$m or [Ar III] 8.991$\mu$m line simultaneously with [Ne II] 12.81$\mu$m, but observing other lines will require exposures with multiple channels.  The brightest [Ne II] 12.81$\mu$m lines from our models, using the fluxes predicted by \citet{Dessart2024} and scaling them with our predicted $\dot{E}_{\rm cool}$ values, will be detectable at $z = 0.1$ with SNR = 10 using an exposure time of $\sim$ 10 ks.  However, if the $L_{\rm PWN}$ and $M_{\rm ej}$ correlations hold beyond the edge of the model grid, then signatures may be detectable at higher redshifts or with lower integration times.

The low spectral peak of the PWN that correlates with strong infrared cooling at late times is difficult to determine from light curve models, but is relevant for later multiwavelength observations.  The low ionization fraction of the ejecta means a low free electron density, so free-free absorption of the radio synchrotron from the PWN should be minimal.  PWNe with low spectral peaks also emit a larger fraction of their energy at radio wavelengths, so these supernovae should be bright in radio on earlier timescales than predicted by previous studies \citep[e.g.][]{Omand2018}.  Getting multiwavelength evidence for the presence of a newborn magnetar is important for excluding scenarios like CSM interaction, which can also produce some higher ionization lines, like [O III].

\section{Conclusion} \label{sec:conc}

This work examines the cooling power from the radiative transfer simulations of magnetar-driven supernovae presented in \citet{Omand2023}, focusing on the characterizing the infrared cooling to better optimize future JWST observations.  Models show little infrared emission at 1 year, a variety of infrared emission profiles at 3 years, and are dominated by infrared emission at 6 years.  The fraction of cooling in the infrared show a sharp dependence on the luminosity of the PWN, being optically-dominated above 10$^{41}$ erg s$^{-1}$, infrared-dominated below 10$^{39}$ erg s$^{-1}$, and showing a variety of properties at intermediate values.  The fraction of cooling in infrared also increases at higher ejecta masses and lower average energy in the PWN spectrum, although this dependence is not as strong.  The cooling luminosity of [Ne II] 12.81$\mu$m increases with increasing PWN luminosity and increasing ejecta mass.  Cooling at 1 year is mostly done through optical O and S lines, with some contribution from Mg and Ca.  Optical coolants are the same at 3 years, but infrared cooling is more pronounced, with contributions from Ar, Ni, and Ne lines.  The most ubiquitous line and strongest infrared coolant at both 3 and 6 years is the [Ne II] 12.81$\mu$m line.  At 6 years, the only strong optical coolant is the [Ca II] $\lambda\lambda$ 7291.5, 7323.9 doublet; while the mid-infrared has the same coolants as at 3 years, and strong S, Fe, Si, and O lines appear in the far-infrared.  For the models with the brightest [Ne II] 12.81$\mu$m line, the line should be detectable out to $z \sim 0.1$ with $\lesssim$ 10 ks exposure time.  Supernovae with high inferred magnetic field strength should transition from optical- to infrared-dominated at an earlier timescale, and infrared-dominated supernovae are good candidates for radio follow-up to detect the PWN.    
\\
\section*{Acknowledgements}

The author thanks Jacob Wise, Steve Schulze, and Gavin Lamb for their helpful discussions, and the anonymous referee for their helpful comments.  C. M. B. O. acknowledges support from the Royal Society (grant Nos. DHF-R1-221175 and DHF-ERE-221005).

\section*{Data Availability}

All data used is publicly available at the link in Footnote \ref{fn:link}.

\bibliography{ref}{}
\bibliographystyle{mnras}

\end{document}